\documentstyle[prl,aps,epsfig,multicol]{revtex}
\input epsf

\begin{document}
\draft
\twocolumn[\hsize\textwidth\columnwidth\hsize\csname@twocolumnfalse\endcsname

\title{Low temperature spin fluctuations in geometrically frustrated 
Yb$_3$Ga$_5$O$_{12}$.}

\author{J.A.Hodges, P.Bonville}
\address{Centre d'Etudes de Saclay, Service de Physique de
l'Etat Condens\'e \\ 91191 Gif-sur-Yvette, France}
\author{M.Rams, K.Kr\'olas}
\address{Institute of Physics, Jagiellonian University \\ 30-059 Krak\'ow, Poland}

\maketitle

\begin{abstract}

In the garnet structure compound Yb$_3$Ga$_5$O$_{12}$, the Yb$^{3+}$ ions
(ground state effective spin S$'$ = 1/2) are situated on two interpenetrating 
corner sharing triangular sublattices such that frustrated magnetic 
interactions are possible. 
Previous specific heat measurements have evidenced the development of short 
range magnetic correlations below $\sim$ 0.5K and a $\lambda$-transition at 
0.054K (Filippi et al. J. Phys. C: Solid State Physics {\bf 13} (1980) 1277). 
From $^{170}$Yb M\"ossbauer spectroscopy measurements down to 36\,mK, we find 
there is no static magnetic order at temperatures below that of the 
$\lambda$-transition. Below $\sim$ 0.3\,K, the fluctuation frequency of the 
short range correlated Yb$^{3+}$ moments progressively slows down and as 
T $\to$ 0, it tends to a quasi-saturated value of $3 \times 10^9$\,s$^{-1}$. 
We also examined the Yb$^{3+}$ paramagnetic relaxation rates up to 300\,K
using $^{172}$Yb perturbed angular correlation measurements: they evidence 
phonon driven processes. 

\end{abstract}

]

\section{Introduction} \label{sectionintro}

For most crystallographically ordered compounds containing magnetic ions, the 
limiting low temperature magnetic ground state involves long range magnetic
order where the spin fluctuations die out as T $\to$ 0. For some particular 
lattice structures however, the geometric arrangement of the magnetic 
ions is such that it may not be possible to simultaneously minimise all 
pairs of interaction energies. The resulting frustration may then lead to a 
situation where long range order does not occur 
\cite{diep,villain}
and where the presence of a large number of low energy states
leads to the continued presence of spin fluctuations as T $\to$ 0.

Systems with frustrated interactions that are of current interest 
include the kagom\'e lattice \cite{broh,wata}, where the ions are arranged on 
a motif of corner sharing triangles, the pyrochlore lattice \cite{moessn},
where the ions are arranged on corner sharing tetrahedra and the garnet 
lattice (R$_3$T$_5$O$_{12}$)
\cite{grenats}, where the rare earths (R) form two interpenetrating, 
non-coplanar, corner sharing triangular sublattices. This geometry 
does allow frustration to be operative provided there is a suitable 
combination of the nature and the size of the rare earth anisotropy and of
the sign of the interionic interactions. A number of the 
rare earth garnets appear to evidence a conventional long range ordered state 
\cite{onn67} suggesting that in these cases, frustration plays a negligible 
role. In fact, to date, frustration has 
been reported to play a major role in only one garnet, Gd$_3$Ga$_5$O$_{12}$ 
\cite{kinney79,hov80,schiffer94,petrenko99,dunsiger00,marshall02}, 
where the S-state Gd$^{3+}$ ion has a very small intrinsic anisotropy and 
where the dominant coupling is antiferromagnetic.

Amongst the garnets made with the non S-state rare earths,  
Yb$_3$Ga$_5$O$_{12}$ is unusual in that the Yb$^{3+}$ ground state shows only
a relatively modest crystal field anisotropy (see below). 
Specific heat data has evidenced a broad peak centred near 0.2\,K, attributed 
to short range correlations and a sharp $\lambda$-peak at 0.054\,K, initially 
attributed to the onset of long range magnetic order
\cite{filippi80}. 
We have carried out $^{170}$Yb M\"ossbauer spectroscopy measurements down to 
0.036\,K in order to examine the behaviour of the Yb$^{3+}$ moments as the 
temperature is lowered through that of the $\lambda$-transition
and to examine the low temperature spin dynamics.
We also report $^{172}$Yb perturbed angular correlation measurements, carried 
out from 14 to 300\,K, which provide information concerning the thermal 
dependence of the Yb$^{3+}$ fluctuation rates in the paramagnetic region.

\section{Background properties}

The single phase polycrystalline sample was prepared by heating the 
constituent oxides to 1100$^\circ$C four times with intermediate grindings.

In the garnet lattice (space group $Ia{\bar 3}d$), the rare earth site point 
symmetry is orthorhombic ($mmm$). The crystal field acts on the Yb$^{3+}$,
$^2F_{7/2}$ state to leave a ground state Kramers doublet which is very well 
isolated from the excited Kramers doublets
\cite{buchan}. 
For Yb$^{3+}$ ions diluted in Y$_3$Ga$_5$O$_{12}$, the ground doublet 
g-values are g$_x$ = 3.73, g$_y$ = 3.60 and g$_z$ = 2.85   
\cite{wolf62,carson60},  and the wave function is derived from
the cubic $\Gamma_7$ state (g = 3.43). In Yb$_3$Ga$_5$O$_{12}$, the
Yb$^{3+}$ g-values should be quite similar.

\begin{figure}
\epsfxsize=6 cm
\centerline{\epsfbox{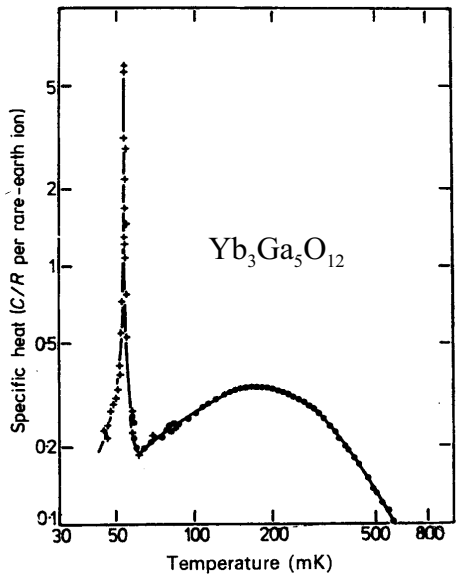}}
\vspace {0.5 cm}
\caption{Specific heat in Yb$_3$Ga$_5$O$_{12}$ reproduced from 
Ref.\protect\cite{filippi80}.}
\label{cp}
\end{figure}

A magnetic 4f-shell contribution to the specific heat is visible below 
$\sim$ 0.5\,K (see Fig.\ref{cp}): there is a broad peak centered near
0.2\,K, followed by a $\lambda$-anomaly at 0.054\,K. The total 
electronic entropy change below 1\,K is very close to the value Rln2 expected
for an isolated Kramers  doublet, and only about 10\% of it is released at 
the $\lambda$-transition 
\cite{filippi80}.
If we attribute the broad peak to exchange driven short range
correlations, then the exchange energy scale is $ \sim$ 0.2\,K.
The susceptibility data \cite{filippi80,ball60}
follow a Curie-Weiss behaviour down to $\sim$ 1.0\,K with a small 
paramagnetic Curie-Weiss temperature ($\theta_p$ = 0.05\,K) 
corresponding to a net ferromagnetic interaction. 
Below $\sim$ 1.0\,K, in the region where the broad specific heat peak occurs
and where the magnetic correlations develop, the thermal dependence of the 
susceptibility falls below
that corresponding to the extrapolated Curie-Weiss dependence. This behaviour
evidences the presence of magnetic correlations which are antiferomagnetic.
In Yb$_3$Ga$_5$O$_{12}$, there is thus evidence for the existence of both
ferromagnetic and antiferromagnetic interactions. The small value for 
$\theta_p$ suggests the two types of interactions have comparable strengths.
The magnetic frustration in Yb$_3$Ga$_5$O$_{12}$ that is evidenced in this 
report, thus appears to be linked to the presence of antiferromagnetic 
correlations within a Heisenberg-like system on triangular sublattices and 
to the presence of interactions with opposite signs. 
The isomorphous compound Gd$_3$Ga$_5$O$_{12}$ where frustration is also
operative, evidences a dominant nearest neighbour interaction which is 
antiferromagnetic and other interactions with competing signs
\cite{kinney79}.

\section{$^{170}$Yb M\"ossbauer measurements} 
\label{sectionmoss}

\subsection{General features} 
\label{submossa}

\begin{figure}
\epsfxsize=6.5 cm
\centerline{\epsfbox{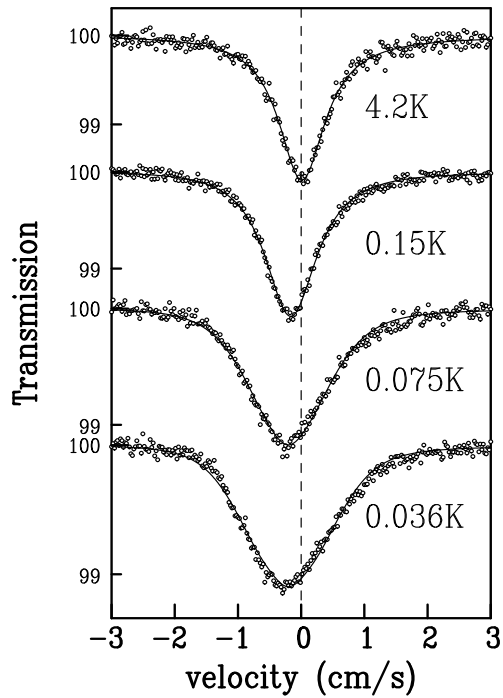}}
\vspace {0.5 cm}
\caption{$^{170}$Yb$^{3+}$ M\"ossbauer absorption in Yb$_3$Ga$_5$O$_{12}$. 
At 4.2\,K, the fitted line 
was obtained using a relaxation lineshape appropriate for paramagnetic
fluctuations. At 0.15, 0.075 and 0.036\,K, the data fits were obtained using
a lineshape appropriate for hyperfine field fluctuations. As the temperature 
decreases, the absorption broadens and its centre of gravity moves towards 
velocities which are more negative than the isomer shift value shown by the 
dashed line.}
\label{spemo}
\end{figure}

The $^{170}$Yb M\"ossbauer absorption measurements (I$_g$ = 0, I$_e$ = 2, 
E$_{\gamma}$ = 84\,keV, 1\,cm/s corresponds to 680\,MHz) were made down to 
0.036\,K in a $^3$He-$^4$He dilution refrigerator using a neutron 
activated TmB$_{12}$ source displaced with a triangular velocity sweep.
Selected spectra at 4.2, 0.15, 0.075 and 0.036\,K are shown in 
Fig.\ref{spemo}. The last two temperatures are situated either side of that of
the specific heat $\lambda$-transition (0.054\,K).

At 4.2\,K, the absorption takes the form of a broad Lorentzian shaped line, 
with a half-width at half maximum, $\Gamma=4$\,mm/s, significantly broader 
than 
the experimental halfwidth of the TmB$_{12}$ source, $\Gamma_0=1.35$\,mm/s. 
As the temperature is lowered to $\sim$ 0.25\,K, the absorption line remains 
Lorentzian shaped and keeps essentially the same width, but it moves slightly
towards negative Doppler velocities. This means the centre of 
gravity of the absorption no longer corresponds to the value of the isomer 
shift, which is close to 0.0\,mm/s relative to the TmB$_{12}$ source. Below 
$\sim$ 0.3\,K, the absorption line moves markedly towards more negative 
velocities, it also progressively broadens and becomes slightly asymmetric
(spectra at 0.075 and 0.036\,K in Fig.\ref{spemo}). 
No resolved hyperfine structure is visible at any temperature, even at
temperatures below that of the $\lambda$-transition. 
This indicates there is no ``static'' long or 
short range magnetic order and that magnetic fluctuations persist down to the 
lowest temperatures. In relation to the characteristic frequency scale 
of the present $^{170}$Yb M\"ossbauer measurements, the absence of a well 
resolved hyperfine structure  means that the fluctuation frequency of the 
Yb$^{3+}$ magnetic moments remains above the threshhold value of 
$\sim$ 3$\times 10^8$\,s$^{-1}$. 
The quantitative analysis of the fluctuation rate is presented in the next
section.

\subsection{Quantitative analysis} 
\label{submossb}

The broad, single-line-nature of the absorption in the paramagnetic region
at 4.2\,K arises because
the Yb$^{3+}$ magnetic hyperfine splitting is ``motional narrowed'' by the 
fast fluctuations of the Yb$^{3+}$ magnetic moment. It is possible to extract 
the fluctuation frequency from the measured lineshape by using a 
paramagnetic spin relaxation model based on a perturbative approach
\cite{imbert}, provided the magnetic hyperfine tensor $\cal A$ is known. 
With the approximation of local axial symmetry, we obtained the components of 
this tensor from a $^{170}$Yb M\"ossbauer measurement on Yb$^{3+}$ ions 
diluted into Y$_3$Ga$_5$O$_{12}$, where because the dilution removes the 
spin-spin coupling, the fully resolved hyperfine splitting is observable. 
We obtained the values:   
$A_z/h$ = 738\,MHz and $A_{\perp}/h$ = 952\,MHz, corresponding respectively
to g-values: g$_z$=2.82 and g$_\perp$=3.63. These values are essentially
equivalent to those previously measured by electron spin resonance
(g$_z$=2.85 and g$_\perp$=(g$_x$+g$_y$)/2=3.66 \cite{ball60}).
We note that with an isotropic ${\cal A}$ tensor and in the fast relaxation 
limit (i.e. when $A/\hbar \ll 1/\tau$, where $1/\tau$ is the Yb$^{3+}$ 
paramagnetic spin relaxation rate), this lineshape model leads to a single
line having a Lorentzian shape with a dynamical half broadening given by:

\begin{equation}
\Delta \Gamma_R = {3 \over 2} (A/\hbar)^2 \tau.
\label{dyn}
\end{equation}

On fitting the data at 4.2\,K using the perturbation approach
\cite{imbert} and  the axially symmetric  $\cal A$ tensor components, we 
obtain the Yb$^{3+}$ paramagnetic spin fluctuation rate: 
$1/\tau \simeq 3.8 \times 10^{10}$\,s$^{-1}$. This rate is constant between 
$\sim$ 0.25\,K and the highest measurement temperature of 80\,K, indicating 
that the driving mechanism is the temperature independent spin-spin coupling
between the Yb$^{3+}$ ions. A rough estimate of the strength of this 
coupling, using the relation: $\hbar/\tau \sim k_B T_{ex}$, yields: 
$T_{ex} \sim 0.3$\,K, consistent with the temperature of the broad maximum of 
the exchange induced specific heat ($\sim$ 0.2\,K).  

The shift of the centre of gravity of the absorption away from the 
temperature independent isomer 
shift value evidenced in Fig.\ref{spemo} is analogous to that previously 
observed at low temperatures in YbAlO$_3$ \cite{bonv78} and in YbBe$_{13}$ 
\cite{bonville86}. It seems to be related to the inadequacy, in these cases, 
of the perturbative relaxation lineshape when the driving mechanism is the 
exchange spin-spin interaction. As discussed in Ref.\onlinecite{bonville86}, 
it is likely that the matrix elements of the  relaxation operator have 
non-vanishing imaginary parts
\cite{afan78,onish78}, and these generate lineshapes whose spectral 
signature is an anomalous shift of the centre of gravity of the absorption.
The fact that the anomalous shift increases as the temperature 
decreases could be linked to the growing influence of the spin-spin 
correlations, hence to the growing inadequacy of the standard perturbative
relaxation lineshape. The relaxation rates obtained from the
perturbation analysis are not influenced by the shift of the centre of gravity
of the absorption since the relaxation rate is linked to the real part of
the matrix elements, i.e. to the width of the absorption line.

At very low temperatures ($T < 0.1$\,K), the shape of the experimental 
absorption can no longer be correctly reproduced by the paramagnetic 
relaxation model. 
Since the specific heat data (Fig.\ref{cp}) show that magnetic correlations 
are present in this temperature region, we fitted the experimental data below 
1\,K using a relaxation model involving hyperfine field fluctuations 
which were considered in the random phase approximation (RPA) 
\cite{dattagupta}. A hyperfine field 
($H_{hf}$) is indeed present at the $^{170}$Yb nucleus when the Yb$^{3+}$ 
moments are short (or long) range correlated  and the size of the field is 
proportional to that of the Yb$^{3+}$ moment. 
The time dependent Hamiltonian for this relaxation lineshape is:

\begin{equation}
{\cal H} = {\cal H}_Q - g_n \mu_n H_{hf} \sum_{j=1,N} I_j f_j(t),
\label{hameff}
\end{equation}

\noindent
where ${\cal H}_Q$ is the quadrupolar hyperfine Hamiltonian, $g_n$ the 
gyromagnetic factor of the excited nuclear state, $\mu_n$ the nuclear Bohr 
magneton, the summation is over the $N$ directions among which the hyperfine
field fluctuates and $f_j(t)$ is a random function of time with appropriate 
values corresponding to the different possible forms of the 
Hamiltonian. For ${\cal H}_Q$, we take the (very small) value we have 
determined from $^{170}$Yb measurements on Yb$^{3+}$ substituted into 
Y$_3$Ga$_5$O$_{12}$. Then, the lineshape depends on a 
single dynamic parameter, the fluctuation frequency of the hyperfine field  
$(1/\tau)_{hf}$, and on the choice of the directions between which the 
fluctuations take place. We obtain very 
satisfactory fits to the data below 0.2\,K (solid lines in Fig.\ref{spemo} at 
0.036, 0.075 and 0.15\,K) by assuming the hyperfine field fluctuates
between the three principal directions of the local coordinate frame, that is,
between  $OX$, $OY$ and $OZ$, the principal directions of the electric 
field gradient tensor.  In the Hamiltonian (\ref{hameff}), this corresponds 
to $N$=6 and $j=\pm X, \pm Y \ {\rm and} \pm Z$. If we assume the 
hyperfine field fluctuates along only one of these axes, we obtain much 
poorer data fits. In the rapid relaxation rate limit, this model yields a 
line of Lorentzian shape with a dynamical half-width:

\begin{equation}
\Delta \Gamma_R = 2 {{(g_n \mu_n H_{hf}/\hbar)^2} \over {(1/\tau)_{hf}}}.
\label{grel}
\end{equation}

The 0.036\,K lineshape is well broadened and this enables us to obtain both 
the magnitude of the hyperfine field and its fluctuation rate.  We find: 
$H_{hf} \simeq 140(10)\,T$, which corresponds to a Yb$^{3+}$ moment of 
$\simeq 1.4$\,$\mu_B$ (for $^{170}$Yb$^{3+}$, 1\,$\mu_B$ yields a hyperfine 
field of 102\,T) and 1/$\tau)_{hf}$ = $3 \times 10^9$\,s$^{-1}$. 
The value for the Yb$^{3+}$ moment is not far from the mean value expected 
both from the average g-tensor (3.4) and from the saturated 
magnetisation measured at 0.09\,K in Ref.\onlinecite{filippi80} 
(1.7\,$\mu_B$).
In the fits of the spectra up to 0.2\,K, we used an intrinsic half-width:
$\Gamma_0$=1.35\,mm/s and we assumed the fluctuating hyperfine field 
has a size which remains constant at the value 140\,T derived 
at 0.036\,K. This assumption should hold as long as the correlations
are well developed, but it cannot be ascertained that it is completely 
correct up to 0.2\,K. 

\begin{figure}
\epsfxsize=8 cm
\centerline{\epsfbox{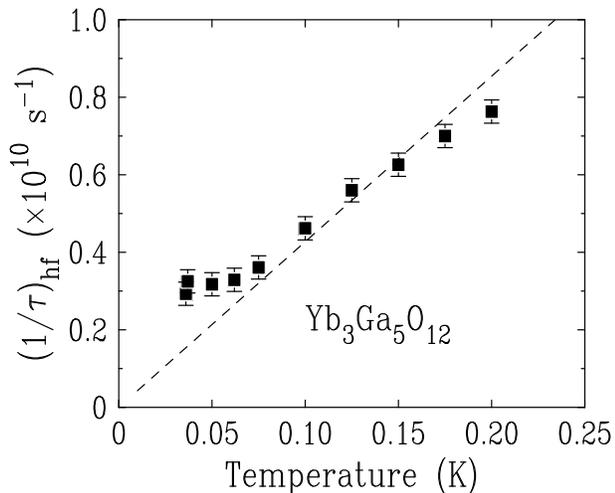}}
\caption{Thermal variations, in Yb$_3$Ga$_5$O$_{12}$, of the Yb$^{3+}$ 
hyperfine field fluctuation frequency 
extracted from the $^{170}$Yb M\"ossbauer spectra.
The dashed line is the law: $\hbar (1/\tau)_{hf} = 0.3 k_BT$.}
\label{nuhpf}
\end{figure}

The thermal variation of $(1/\tau)_{hf}$ is shown on Fig.\ref{nuhpf}. As the 
temperature is lowered over the range 0.2 to 0.1\,K, the 
frequency decreases approximately linearly with a law: 
$\hbar (1/\tau)_{hf} = 0.3 k_B T$, and then below about 0.1\,K, it tends to 
saturate towards the value $3 \times 10^9$\,s$^{-1}$. 
There is essentially no difference between the rates either side of the
specific heat $\lambda$-transition (0.054\,K).
The $T$-linear dependence of $(1/\tau)_{hf}$ only pertains to a 
limited temperature range, and it must be kept in mind that it somehow depends
on the validity of the assumption about the constant magnitude of the
fluctuating hyperfine field. 
We note that a linear variation has also been encountered theoretically
for the case of a frustrated Heisenberg pyrochlore antiferromagnet 
\cite{moessn}.
The decrease in the 
relaxation rate followed by a saturation is similar to the behaviour 
observed (by $\mu$SR \cite{marshall02}) in the isomorphous compound 
Gd$_3$Ga$_5$O$_{12}$.

\section{$^{172}$Yb perturbed angular correlation (PAC) measurements}
\label{sectionpac}

The PAC measurements provide the Yb$^{3+}$ paramagnetic relaxation rates in 
the temperature range 14 - 300\,K. The measurements were made 
using the 91-1094\,keV $\gamma-\gamma$ cascade from the
$^{172}$Lu $\to$ $^{172}$Yb $\beta$ decay. The $^{172}$Lu nuclei are obtained
by proton irradiating the sample, which is then annealed at 800$^\circ$C to
remove irradiation defects. The intermediate level of the cascade, namely the
1172\,keV nuclear excited level of $^{172}$Yb with spin $I=3$ and half-life
8.3\,ns, is used to observe the perturbation of the $\gamma-\gamma$ 
directional correlations due to the hyperfine interactions.
In the case of a static quadrupolar or magnetic hyperfine interaction, 
oscillations are observed in the time evolution of the perturbation factor 
$R(t)$ \cite{frau}.

\begin{figure}
\epsfxsize=7 cm
\centerline{\epsfbox{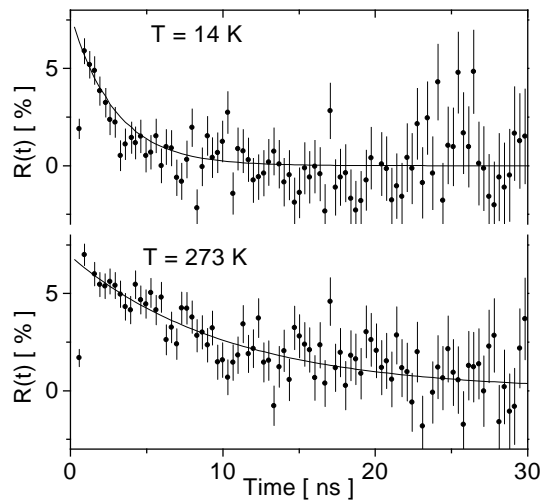}}
\vspace {0.5 cm}
\caption{$^{172}$Yb PAC spectra at 14\,K and 300\,K, fitted to an exponential
decay. The first channel, spoilt due to prompt coincidences, was removed
from the fits.}
\label{pacsp}
\end{figure}

Up to $\sim$300\,K, we observe the perturbation factor $R(t)$ does 
not show oscillations but it is rather an exponential function of time: 
$R(t)=A \exp(-\mu t)$ (see Fig.\ref{pacsp}).
This suggests a dynamic hyperfine interaction.
Paramagnetic relaxation within the ground Yb$^{3+}$ doublet, with effective
spin 1/2, leads to such an exponential decay, in the fast relaxation limit:
$A^{172}/\hbar \ll 1/\tau$, where $A^{172}$ is the hyperfine 
constant of the intermediate $I=3$ level of $^{172}$Yb and $1/\tau$ 
the electronic spin fluctuation frequency. 
Then, in a manner analogous to Eqn.\ref{dyn}, the damping rate 
$\mu$ is given by \cite{abr}:

\begin{equation}
\mu = {3 \over 2} (A^{172}/\hbar)^2 \tau.
\label{dynmu}
\end{equation}

\begin{figure}
\epsfxsize=8 cm
\centerline{\epsfbox{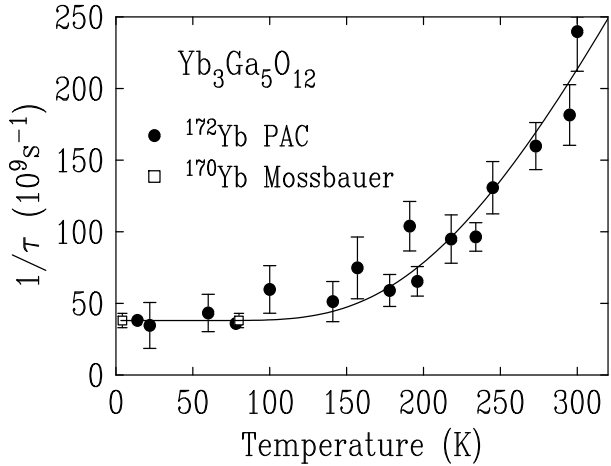}}
\vspace {0.5 cm}
\caption{Thermal variation of the Yb$^{3+}$ 4f shell magnetic fluctuation 
rate in Yb$_3$Ga$_5$O$_{12}$ derived from perturbed angular correlations 
measurements (black dots). Two values obtained from the $^{170}$Yb 
M\"ossbauer spectra at 4.2 and 80\,K are also 
shown (open squares). The solid line is a fit to a sum of spin-spin
and spin-phonon driven relaxation rates (see text).}
\label{nupac}
\end{figure}

The value for the hyperfine constant of the $I=3$ level of $^{172}$Yb can be 
obtained by scaling the mean $A^{170}$ value appropriate for the Yb$^{3+}$ 
ground state ($A^{170}/h \simeq 884$\,MHz) with the nuclear
g-factors. We obtain: $A^{172}/h \simeq 565$\,MHz. 
The thermal variation of $1/\tau$ derived from that of 
$\mu$ is shown in Fig.\ref{nupac} which also shows the $^{170}$Yb M\"ossbauer 
derived values at 4.2 and 80\,K. The two sets of values agree quite well. 
The relaxation rate is constant up to $\sim$ 150\,K, then increases 
monotonically. The observed thermal dependence can be fitted to the sum of a 
temperature independent spin-spin term and an exponential term associated with
a two-phonon real process through the excited crystal field states
(Orbach process) \cite{orbach}:
\begin{equation} 
{1 \over \tau} = ({1 \over \tau})_{ss} + B\  \exp(-{\Delta \over {k_BT}}),
\label{orb}
\end{equation}
where $\Delta$ is the energy of an excited crystal field level. The fit
yields: $(1/\tau)_{ss} = 3.8 \times 10^{10}$\,s$^{-1}$, $B=3.3 \times 10^{12}$
\,s$^{-1}$ and $\Delta$ = 880(50)\,K. This last value is in very good 
agreement with the mean energy distance between the Yb$^{3+}$ ground doublet 
and the three closely spaced excited crystal field doublets ($\sim$ 850\,K)
\cite{buchan}.
Additional PAC measurements concerning the thermal variation of the electric 
field gradient at the $^{172}$Yb nucleus above 300\,K are reported in 
Ref.\onlinecite{krolas}.

\section{Discussion and conclusions}

Previous specific heat measurements 
\cite{filippi80} have shown that, in Yb$_3$Ga$_5$O$_{12}$, magnetic 
correlations develop below $\sim$ 0.5\,K and that a $\lambda$-transition
occurs at 0.054\,K. Our present investigations, using $^{170}$Yb M\"ossbauer
spectroscopy down to 36\,mK and
$^{172}$Yb perturbed angular correlations measurements up to 300\,K, provide 
insight into the correlations and the dynamic behaviour of the Yb$^{3+}$ spins
over a wide temperature range.

Above $\sim$ 0.5\,K, the Yb$^{3+}$ magnetic moments undergo paramagnetic
fluctuations. Up to $\sim$ 150\,K, the fluctuation rate has a temperature
independent value of $\simeq 3.8 \times 10^{10}$\,s$^{-1}$
and the driving mechanism is the exchange interaction between the the 
Yb$^{3+}$ spins. Above 150\,K, additional temperature 
dependent relaxation occurs through coupling to phonons according to a 
two-phonon Orbach process involving the excited crystal field states near 
850\,K.

The low temperature Yb$^{3+}$ magnetic correlations show up in the $^{170}$Yb 
M\"ossbauer measurements below $\sim$ 0.3\,K through changes in the lineshape.
The fluctuation frequency of the correlated moments decreases as the 
temperature is lowered and below 0.1\,K, it tends to a quasi-saturated value
of 3$\times 10^9$\,s$^{-1}$. This is a quite high value and in fact, 
Yb$_3$Ga$_5$O$_{12}$ is the only known compound where the T $\to$ 0 
spin fluctuation rate is rapid enough to fall within the $^{170}$Yb 
M\"ossbauer spectroscopy frequency window. 
The T $\to$ 0 state in Yb$_3$Ga$_5$O$_{12}$ is therefore a dynamic 
short range correlated spin-liquid state. 
On crossing the temperatures of the specific heat $\lambda$-anomaly,
there is no significant change in the $^{170}$Yb M\"ossbauer lineshape.
Usually, a transition to a long range magnetically ordered phase reveals 
itself in the M\"ossbauer spectra by the appearance of a well defined 
magnetic hyperfine splitting. This does not appear in the present case,  
clearly showing there is no long range order at temperatures below that of 
the specific heat peak. This peak has the intriguing characteristic that the 
associated entropy gain is very small, i.e. it amounts to about 10\% of the 
total Rln2 entropy gain associated with the Yb$^{3+}$ ground state Kramers 
doublet. 

Some examples are already known of frustrated systems where the specific heat 
peak has an associated low entropy. 
In the pyrochlore compound Gd$_2$Ti$_2$O$_7$, there is an entropy gain of 
50\% of $R\ln8$ at the transition at 1\,K \cite{raju},
which has been shown by neutron diffraction to involve long 
range magnetic order \cite{champ}.
Yb$_2$Ti$_2$O$_7$ presents an entropy gain of about 20\% of $R\ln2$ 
\cite{blote} 
at the transition at 0.25\,K which has been shown to be associated with a 
first order change in the fluctuation rate of the correlated Yb$^{3+}$ 
moments 
\cite{hodges}.
In Yb$_3$Ga$_5$O$_{12}$, the reduced size specific heat peak is linked neither
to the appearance of long range order nor to a mesurable change in the 
fluctuation rate of the short range correlated moments. 
In fact, we observe a slowing down 
of the fluctuations of the correlated moments below $\sim$ 0.3\,K, i.e. at
temperatures well above that of the specific heat peak, and the fluctuation 
rate is quasi temperature independent in the region where the 
specific heat peak occurs.

Usually for frustrated systems presenting a phase transition, the
temperature at which this transition occurs is sizeably lower than the 
temperature associated with the strength of the interionic interaction. 
The energy scale of this interaction may be estimated in different ways: from
the paramagnetic Curie-Weiss temperature, from the temperature where the 
specific heat evidences a magnetic contribution (in the present case, the 
temperature of the broad maximum) and from the paramagnetic 
spin-spin relaxation rate. In Yb$_3$Ga$_5$O$_{12}$, the last two methods lead 
to an interaction of equivalent strength 0.2 - 0.3\,K, whereas the 
paramagnetic Curie temperature ($\theta_p$), is smaller (0.05\,K). The 
correspondence between $\theta_p$ and the temperature of the specific heat 
peak appears to be a mere coincidence. The precise origin of this low 
entropy specific heat peak within a spin liquid phase remains an 
unresolved issue.

\medskip

{\bf Acknowledgments.}

We thank Mrs Anne Forget for preparing the samples and Mrs. J. Wojtkowska of 
the Institute for Nuclear Studies, Otwock, Poland for the proton irradiation 
for the PAC experiments.

\end{document}